\documentclass[fleqn=-1in,11pt,twocolumn]{wlscirep}
\DeclareUnicodeCharacter{2062}{}
\usepackage[normalem]{ulem}
\usepackage{lmodern}

\usepackage{authblk}
\usepackage[utf8]{inputenc}
\usepackage[T1]{fontenc}
\usepackage{mathrsfs}
\usepackage{textcomp}
\usepackage{units}
\usepackage{array,mathtools}
\usepackage{stmaryrd}
\usepackage{scrextend}
\usepackage{nicefrac}
\usepackage{multirow}
\usepackage{makecell}
\usepackage{bm}
\newcolumntype{C}{>{$}c<{$}}
\AtBeginDocument{
\heavyrulewidth=.08em
\lightrulewidth=.05em
\cmidrulewidth=.03em
\belowrulesep=.65ex
\belowbottomsep=0pt
\aboverulesep=.4ex
\cmidrulesep=\doublerulesep
\cmidrulekern=.5em
\defaultaddspace=.5em
}
\usepackage{booktabs}
\usepackage[caption=false]{subfig}
\usepackage{graphicx}
\usepackage{dcolumn}
\usepackage{bm}
\usepackage{xspace}
\usepackage{cuted}

\newcommand{\s}{\sigma}
\newcommand{\vop}{\hat{{V}}}
\newcommand{\hop}{\hat{H}}
\newcommand{\mM}{\bm{m}}
\newcommand{\bq}{\bm{q}}
\newcommand{\bk}{\bm{k}}
\newcommand{\bB}{\bm{B}}
\newcommand{\mN}{N}
\newcommand{\mF}{\nu}

\newcommand{\mU}{u}
\newcommand{\trev}{\hat{\mathcal{T}}}
\newcommand{\liou}{\hat{\mathcal{L}}}
\newcommand{\der}[2]{\frac{d#1}{d B_{\omega\bq}^{#2}}}
\newcommand{\derm}[2]{\frac{d#1}{d B_{-\omega-\bq}^#2}}
\newcommand{\pop}{\hat{P}}
\newcommand\redsout{\bgroup\markoverwith{\textcolor{mycolor}{\rule[0.5ex]{2pt}{0.4pt}}}\ULon}

\definecolor{mycolor}{rgb}{0, 0, 0}   
 
\def\QE{\textsc{Quantum ESPRESSO}\,}

\newcommand{\editor}[2]{%
  \expandafter\newcommand\csname #1note\endcsname[1]{%
    \textcolor{#2}{(\textbf{#1:} ##1)}}%
  \expandafter\newcommand\csname #1\endcsname[1]{%
    \textcolor{#2}{##1}}%
  \expandafter\newcommand\csname #1cancel\endcsname[1]{%
    \textcolor{#2}{\sout{##1}}}%
  \expandafter\newcommand\csname #1change\endcsname[2]{%
    \textcolor{#2}{\sout{##1} ##2}}%
  \newenvironment{#1text}{\color{#2}}{\color{black}}
}

\editor{LB}{blue}
\editor{IT}{red}

\title{Magnons from time-dependent density-functional perturbation theory and nonempirical Hubbard functionals}

\author[1,$\dagger$]{Luca Binci\thanks{Present address: Department of Materials Science \& Engineering, University of California Berkeley, Berkeley, CA, 94720, USA; Materials Sciences Division, Lawrence Berkeley National Laboratory, Berkeley, CA, 94720, USA}}
\author[1,2]{Nicola Marzari}
\author[2,+]{Iurii Timrov}

\affil[1]{Theory and Simulation of Materials (THEOS), and National Centre for Computational Design and Discovery of Novel Materials (MARVEL), École Polytechnique Fédérale de Lausanne, CH-1015 Lausanne, Switzerland}
\affil[2]{PSI Center for Scientific Computing,
Theory, and Data, 5232 Villigen PSI, Switzerland}

\affil[$\dagger$]{Email: lbinci@berkeley.edu}
\affil[+]{Email: iurii.timrov@psi.ch}

\begin{abstract}
\normalfont
Spin excitations play a fundamental role in understanding magnetic properties of materials, and have significant technological implications for magnonic devices. However, accurately modeling these in transition-metal and rare-earth compounds remains a formidable challenge. 
Here, we present a fully first-principles approach for calculating spin-wave spectra based on time-dependent (TD) density-functional perturbation theory (DFPT), using nonempirical Hubbard functionals. This approach is implemented in a general noncollinear formulation, enabling the study of magnons in both collinear and noncollinear magnetic systems.
Unlike methods that rely on empirical Hubbard $U$ parameters to describe the ground state, and Heisenberg Hamiltonians for describing magnetic excitations, the methodology developed here probes directly the dynamical spin susceptibility (efficiently evaluated with TDDFPT throught the Liouville-Lanczos approach), and treats the linear variation of the Hubbard augmentation (in itself calculated non-empirically) in full at a self-consistent level. Furthermore, the method satisfies the Goldstone condition without requiring empirical rescaling of the exchange-correlation kernel or explicit enforcement of sum rules, in contrast to existing state-of-the-art techniques. We benchmark the novel computational scheme on prototypical transition-metal monoxides NiO and MnO, showing remarkable agreement with experiments and highlighting the fundamental role of these newly implemented Hubbard corrections. The method holds great promise for describing collective spin excitations in complex materials containing localized electronic states.
\end{abstract}

\begin{document}

\flushbottom 
\maketitle

\thispagestyle{empty}
\section*{INTRODUCTION}

In recent years, several intriguing research directions in spin excitations have attracted much attention; notably, these include magnons in two-dimensional materials~\cite{avsar_colloquium_2020,gorni_first-principles_2023,ke_electron_2021} and in altermagnets~\cite{Smejkal:2023, Cui:2023, Leenders:2024, Costa:2024}, and the coupling of magnons with other quasiparticles like excitons~\cite{olsen_unified_2021,bae_exciton-coupled_2022}, phonons~\cite{bistoni_intrinsic_2021, bonini_frequency_2023, delugas_magnon-phonon_2023, ren_adiabatic_2024}, and plasmons~\cite{Ghosh:2023, Dyrdal:2023, Falch:2024}. These phenomena often occur in complex materials containing magnetic transition-metal and/or rare-earth ions, and characterized as Mott-Hubbard or charge-transfer insulators~\cite{Imada:1998}. While there have been many experimental breakthroughs in studying collective spin excitations~\cite{Toth:2016, Pershoguba:2018, Chen:2018, Scheie:2022, Huang:2024}, theoretical and computational investigations remain challenging. To model the ground state of this class of materials, density-functional theory (DFT)~\cite{hohenberg_inhomogeneous_1964, kohn_self-consistent_1965} 
is typically used, 
where particular attention must be given to the selection of the exchange-correlation (xc) functional, as it strongly influences the accuracy of the results. While standard local spin-density approximation (LSDA) and spin-polarized generalized-gradient approximation ($\sigma$-GGA) provide satisfactory results for itinerant magnetic metals, they are inaccurate for insulating transition-metal and rare-earth compounds due to strong self-interaction errors (SIEs) for partially filled and localized $d$ and $f$ electrons~\cite{Perdew:1981, MoriSanchez:2006}. To address these challenges, more advanced functionals have been developed, among which Hubbard-corrected DFT functionals ($\mathrm{DFT}+U$)~\cite{anisimov:1991, Liechtenstein:1995, Dudarev:1998} stand out for their capability to correct SIEs~\cite{Kulik:2006, Kulik:2008} and low computational cost. The value of Hubbard $U$ is critical, and empirical tuning of $U$ based on experimental results is a popular strategy; still, not only it requires accurate reference data, which are not always available, but its tranferability to properties that are often not fitted is debatable. To overcome these limitations, several first-principles approaches to compute $U$ have been developed, including constrained DFT (cDFT)~\cite{Dederichs:1984, Mcmahan:1988, Gunnarsson:1989, Hybertsen:1989, Gunnarsson:1990, Pickett:1998, 
Shishkin:2016}, Hartree-Fock-based methods~\cite{Mosey:2007, Mosey:2008, Andriotis:2010, Agapito:2015, TancogneDejean:2020, Lee:2020}, and the constrained random phase approximation (cRPA)~\cite{Springer:1998, Kotani:2000, Aryasetiawan:2004, Aryasetiawan:2006}. Machine learning techniques for determining Hubbard parameters have also emerged in recent years~\cite{Yu:2020, Yu:2023, Cai:2024, Uhrin:2024, Das:2024}, which provide a fast and attractive route. The linear-response formulation of cDFT~\cite{cococcioni_linear_2005} has gained widespread popularity due to its simplicity and accuracy, and its recent reformulation using density-functional perturbation theory (DFPT)~\cite{timrov_hubbard_2018, timrov_self-consistent_2021,binci_noncollinear_2023} has further broadened its success. The physical rationale behind the linear-response determination of $U$ relies on the heuristical imposition of piecewise linearity of the total energy of the system as a function of the occupation of the target Hubbard manifold~\cite{cococcioni_linear_2005}. Investigations of various magnetic materials using $U$ from DFPT have proven to be accurate and effective~\cite{ricca_self-consistent_2019, floris_hubbard-corrected_2020, zhou_ab_2021, Grassano2023, Bonfa:2024, Macke2024, Ponet2024, Gelin:2024, Chang:2025}, making it an appealing approach for the description of spin waves. 

The theoretical modeling of spin waves (magnons) can in general be achieved using a wide array of methodologies. One of the most popular techniques involves model spin Hamiltonians, particularly the Heisenberg model, which relies on the adiabatic assumption that the time scales of magnons and electrons differ enough to allow the local electronic structure to adapt to the presence of magnons. The Heisenberg Hamiltonian is parametrized with interatomic exchange interactions $J$ and other magnetic interaction parameters; e.g., single-ion anisotropy and/or Dzyaloshinskii-Moriya (DM) interactions~\cite{szilva_quantitative_2023}. These parameters are often obtained empirically by fitting them to experimental magnon dispersions. From a theoretical perspective, they can also be calculated from first principles using established approaches such as total energy differences~\cite{sabani_ab_2020}, spin-spiral energy dispersions based on the generalized Bloch theorem~\cite{halilov_adiabatic_1998, gebauer_magnons_2000,jacobsson_exchange_2013}, and the infinitesimal-rotations method based on the magnetic force theorem~\cite{liechtenstein_local_1987, solovyev_crucial_1996, pajda_ab_2001, korotin_calculation_2015}. 
For instance, in the prototypical transition-metal oxides NiO and MnO, the $J$ parameters have been computed using some of the aforementioned methods on top of DFT calculations using different xc functionals and corrective methods to describe the ground state. These approaches include self-interaction correction methods~\cite{Kodderitzsch:2002, fischer_exchange_2009, Archer:2011}, hybrid functionals~\cite{Archer:2011}, and DFT+$U$ with either empirical or \textit{ab initio} $U$ values~\cite{jacobsson_exchange_2013, Solovyev:1998, Gopal:2017}. However, the $J$ parameters exhibit strong sensitivity to the value of the Hubbard $U$, making comparisons across different studies and with experimental results challenging.

Once the Heisenberg Hamiltonian parametrization is set, magnon dispersions can be determined using linear spin-wave theory (LSWT)~\cite{Bloch:1930, Slater:1930, Holstein:1940}. 
Mapping experimental magnon dispersions to lattice spin models has proven to be very effective, offering a valuable tool for investigating complex systems, particularly magnetic surfaces~\cite{Zakeri2012} and skyrmions~\cite{Heinze2011}. However, using Heisenberg Hamiltonians often requires prior knowledge of the specific magnetic interactions to include, which can be problematic if no experimental data are available. Moreover, Heisenberg Hamiltonians fail to account for the effects stemming from the low-frequency Stoner excitations of metals, which lead to Landau damping~\cite{Landau:1946} of magnons, and the application of the Heisenberg model to itinerant metallic magnets is in itself questionable. Additionally, in complex systems, the large number of $J$ parameters and weak DM interactions can introduce intricacies in the determination of these quantities. Furthermore, modeling chiral magnons in altermagnets using Heisenberg Hamiltonians is particularly challenging, as it demands high-resolution accuracy for anisotropic splittings of exchange interactions between distant neighbors, such as the splitting of $J$ parameters for the 7th nearest neighbors in MnF$_2$~\cite{Morano:2024} or the 10th nearest neighbors in MnTe~\cite{Liu:2024}. Therefore, a first-principles approach for modeling magnons that is independent of underlying spin models is highly desirable.

An alternative approach for modeling spin waves is to calculate directly the electronic response to an external magnetic field perturbation, by evaluating explicitly the spin-spin susceptibility tensor. Two popular strategies to tackle this task are time-dependent density-functional theory (TDDFT)~\cite{gross_local_1985} and many-body perturbation theory (MBPT)~\cite{Hedin:1965}. TDDFT equations are typically solved in the linear-response regime in the frequency domain, assuming a small external magnetic field perturbation, and can be addressed using Dyson~\cite{rousseau_efficient_2012, Lounis:2010, Lounis:2011, Buczek:2011b, dosSantosDias:2015, Wysocki:2017, Singh:2019, skovhus_dynamic_2021,Sandratskii2012,Odashima2013}, Sternheimer~\cite{savrasov_linear_1998, cao_ab_2018, liu_implementation_2023}, or Liouville-Lanczos (LL)~\cite{gorni_spin_2018} approaches. The strong perturbation regime can be accessed by solving TDDFT equations using real-time propagation, which enables the modeling of ultrafast phenomena~\cite{tancogne-dejean_time-dependent_2020}. MBPT techniques, consisting in the solution of the Bethe-Salpeter equation on top of the LSDA or $GW$ ground state, have been applied for modeling magnons~\cite{karlsson_spin-wave_2000, sasioglu_wannier-function_2010, muller_electron-magnon_2019, olsen_unified_2021}. 
While MBPT is known to provide a more accurate description of absorption spectra of solids than TDDFT with adiabatic LSDA (ALSDA), it involves higher computational costs. However, for magnons a comprehensive comparison of the accuracy and computational costs of MBPT and TDDFT across a wide range of materials has not yet been conducted (previous studies have primarily focused on elementary ferromagnets such as Fe, Ni, and Co). Additionally, most MBPT and TDDFT implementations suffer from violations of the Goldstone condition (see, e.g., Refs.~\citenum{skovhus_dynamic_2021, Muller:2016}).
Improved versions of TDDFT embodying more advanced xc functionals are attractive because they could deliver accurate magnon predictions while preserving moderate computational costs. In this context, extending TDDFT to incorporate Hubbard $U$ corrections has proven effective for absorption spectroscopy~\cite{lee_dynamical_2010, orhan_tddft_2019, tancogne-dejean_self-consistent_2017}. However, this extension~\cite{Lounisnote} to the case of magnons has only been explored in a few works, all using empirical $U$: one when solving the Sternheimer equation with a finite-difference scheme~\cite{liu_implementation_2023}, and another by solving the Dyson equation with an additional empirical rescaling of the xc kernel to enforce the Goldstone condition~\cite{Skovhus:2022, Skovhus:2022b}, making these approaches not being entirely \textit{ab initio}. Instead, a fully first-principles Hubbard approach is highly desirable, with $U$ computed using one of the aforementioned methods and treated self-consistently and efficiently when solving the TDDFT equations, also satisfying the Goldstone theorem.

Here, we propose a first-principles methodology for the evaluation of spin-fluctuation spectra which is based on time-dependent density-functional perturbation theory (TDDFPT)~\cite{gross_local_1985, Baroni:2012} with nonempirical Hubbard functionals \cite{binci_noncollinear_2023} in the LL scheme \cite{walker_efficient_2006} and in a general noncollinear formulation. The key features of the novel approach are: (i)~the Hubbard $U$ parameter is not arbitrarily adjusted, but is calculated using DFPT \cite{cococcioni_linear_2005,baroni_phonons_2001,timrov_hubbard_2018} simultaneously optimizing also the crystal structure in a well-defined self-consistent protocol \cite{cococcioni_energetics_2019, timrov_self-consistent_2021}, (ii)~the approach probes explicitly the dynamical spin susceptibility, that is linked to the experimentally measurable double-differential cross section, thus providing a direct comparison with experiments~\cite{gorni_spin_2018}, (iii)~the approach is implemented in a general noncollinear formulation, thus allowing to study systems with collinear and noncollinear ground states, and (iv)~the Goldstone theorem is satisfied without the need in empirical rescaling of the xc kernel~\cite{Skovhus:2022, Skovhus:2022b} or explicit enforcement of sum rules~\cite{Lounis:2010, Lounis:2011}, in contrast to methods based on the solution of the Dyson equation. Besides avoiding any reference to the electronic empty states, as routinely done in DFPT \cite{baroni_greens-function_1987, baroni_phonons_2001}, the greatest advantage of the LL approach for step (ii) relies in the fact that a single linear-response calculation enables the evaluation of a column of the spin susceptibility $\chi_{\alpha\alpha'}(\bq,\omega)$ at a large number frequencies in an inexpensive way as a postprocessing step. This is a desirable feature -- especially when the magnon spectrum of a material is not known \textit{a priori} -- because it allows a facile identification of the spin excitations along the frequency axis, without the need of scanning several values of $\omega$, which instead would be required in the Sternheimer and Dyson approaches~\cite{savrasov_linear_1998,rousseau_efficient_2012}. The LL method was already successfully applied to optical absorption spectroscopy \cite{walker_efficient_2006, rocca_turbo_2008}, electron-energy loss spectroscopy~\cite{timrov_electron_2013}, and inelastic neutron scattering spectroscopy~\cite{gorni_spin_2018} (although limited to standard xc functionals), and it includes the self-consistent readjustment of the charge and magnetization densities. In this work, we explicitly account for the first-order variation of the Hubbard potential, similarly to the static phonon DFPT implementation \cite{floris_vibrational_2011,floris_hubbard-corrected_2020}. However, in the present noncollinear dynamical case we demonstrate that the linear response equation that is antiresonant with the frequency $\omega$ exhibits a reversal of the Hubbard magnetization -- i.e. the magnetization projected onto the localized ($3d$ or $4f$) target manifold. Ultimately, the method proposed here works directly with the linearization of the Kohn-Sham (KS) Bloch states in reciprocal space, and thus it bypasses intermediate post-processing steps like Wannierizations~\cite{korotin_calculation_2015,Xiangzhou2020,rousseau_efficient_2012}, improving user-friendliness and automation. 

We apply this novel approach to the study of the magnon dispersions of NiO and MnO. By consistently addressing the electronic, structural, and magnetic degrees of freedom, we achieve a highly accurate determination of spin waves, comparable to results from other advanced methods like DMFT \cite{wan_calculation_2006} or $GW$ \cite{kotani_spin_2008,ke_electron_2021} but at more moderate computational cost. As a byproduct of the calculations, we fit the magnon dispersions in order to extract the magnetic exchange parameters, which are then used to explain features of the magnon dispersions, and link them to the different magnitude of the rhombohedral distortions detected in the two materials investigated. These structural distortions are shown to be correctly captured by the present approach, thanks to the use of an iterative evaluation of the Hubbard parameters and structural optimizations, yielding the self-consistent Hubbard $U$ and crystal structure.

\section*{RESULTS} 

\subsection*{Time-dependent density-functional perturbation theory with Hubbard corrections}

TDDFPT is a dynamical generalization of static DFPT~\cite{baroni_phonons_2001, gonze_dynamical_1997}, where the external perturbation is decomposed into monochromatic dynamical components characterized by wave-vectors $\bq$ and frequency $\omega$ ~\cite{gorni_spin_2018}. Due to the frequency dependence, two first-order (Sternheimer) equations must be solved: one \emph{resonant} with the perturbation frequency $\omega$ at wave-vector $\bq$, and one \emph{antiresonant} with $-\omega$ at wave-vector $-\bq$. Different methods have been developed to handle these equations. For example, in Ref.~\citenum{cao_ab_2018} the antiresonant equation was explicitly solved at $(-\bq,-\omega)$, while Ref.~\citenum{gorni_spin_2018} applied the time-reversal operator $\trev=\iota\sigma_y \hat{K}$ (with $\hat{K}$ being the complex-conjugate operator) to the antiresonat equation, which restores the positive sign of $\omega$ and $\bq$ in the response quantities while reversing the sign of the magnetic xc potential $\bm{B}_\mathrm{xc}$. The operator $\trev$ has also been used in static DFPT to calculate phonons~\cite{urru_density_2019} and the Hubbard $U$ parameter~\cite{binci_noncollinear_2023}; given its formal elegance, in this work we follow this methodology. 

The core of this paper is an extension of the TDDFPT formalism from Ref.~\citenum{gorni_spin_2018} to noncollinear Hubbard functionals~\cite{binci_noncollinear_2023}. 
We present the general formalism for both metals and insulators using norm-conserving pseudopotentials, and we use Hartree atomic units. It is worth noting that extending the current formalism to ultrasoft pseudopotentials~\cite{vanderbilt_soft_1990} and the projector augmented wave (PAW) method~\cite{blochl_projector_1994} is straightforward but rather involved~\cite{walker_ultrasoft_2007, motornyi_electron_2020}, and could be addressed in future work.

\subsubsection*{Ground state}

In a noncollinear DFT+$U$ scheme, the Hubbard occupation is a $2\times2$ matrix in spin space~\cite{dudarev_parametrization_2019, tancogne-dejean_self-consistent_2017, savrasov_ground_2000}. The implementation of this formalism supports L\"owdin-orthogonalized pseudo-atomic orbitals, which define the localized Hubbard subspace of interest~\cite{binci_noncollinear_2023}. For spinorial quantities, we adopt the notation $|\Psi_i\rangle=\sum_{\s}|\psi^\s_i,\s\rangle$ and $|\Phi^I_m\rangle=\sum_{\s}|\phi^{I\s}_m,\s\rangle$ for the KS spinor and the Hubbard atomic states, respectively (see Ref.~\citenum{binci_noncollinear_2023} for more detailed definitions). 
Here, $i$ is the collective index for quasimomentum and KS band indices $i=n\bk$, $I$ is the atomic site index, $m$ is the magnetic quantum number, and $\sigma$ is the spin index.
The Hubbard occupation matrix is defined as $N_{mm'}^{I}=\sum_i \tilde{\theta}_i \langle\Psi_i|\pop^{I}_{m'm}|\Psi_i\rangle$, where 
$\tilde{\theta}_i$ are the electronic occupancies -- which equal to 0 and 1 for empty and occupied states at zero temperature, respectively, and have intermediate values for metals around the Fermi level -- and
$\pop^{I}_{m'm}=|\Phi_{m'}^{I}\rangle\langle\Phi^I_m|$ is the projector on the Hubbard subspace. In spin-resolved components, the Hubbard occupation matrix reads:
\begin{equation}
    \mN_{mm'}^{I\s\s'} = \sum_i \Tilde{\theta}_i\,\big\langle\psi_i^{\s'}\big|\phi^{I}_{m'}\big\rangle\big\langle\phi^{I}_{m}\big|\psi_i^{\s}\big\rangle.
    \label{eq:occ_matrix}
\end{equation}
In terms of these quantities, the Hubbard energy is given by~\cite{binci_noncollinear_2023}:
\begin{equation}
    E_U = \sum_{Im} \frac{U^I}{2} \mathrm{Tr}\,\Big( \mN^I_{mm}-\sum_{m'}\mN^I_{mm'}\mN^I_{m'm}\Big),
\end{equation}
where the trace ($\mathrm{Tr}$) is taken over the spin degrees of freedom.

As was mentioned earlier, in order to exploit the time-reversal operator $\trev$ in the antiresonant Sternheimer equation, we need to determine how $N^{I}_{mm'}$ and $\pop^{I}_{mm'}$ transform under time reversal. To this aim, we use the completeness property~\cite{sakurai_modern_2020} of the Pauli matrices $\bm{\s} = (\sigma^x, \sigma^y, \sigma^z)$: $\bm{\s}_{\zeta\zeta'}\cdot\bm{\s}_{\xi\xi'}=2\delta_{\zeta\xi'}\delta_{\zeta'\xi}-\delta_{\zeta\zeta'}\delta_{\xi'\xi}$, which allows to rewrite the occupation matrix~\eqref{eq:occ_matrix} as~\cite{binci_noncollinear_2023}:
\begin{equation}
    \big(\mN_{mm'}^{I[\mM]}\big)^{\s\s'}=\frac{1}{2}\,\Big(n_{mm'}^{I}\,\delta^{\s\s'} + \bm{m}^{I}_{mm'}\cdot \bm{\s}^{\s\s'}\Big),
    \label{eq:occ_matrix2}
\end{equation}
where $n^{I}_{mm'}=\mathrm{Tr} \,\big[\mN^{I}_{mm'}\big]$ and $\bm{m}^{I}_{mm'} = \sum_{\s\s'}\mN^{I\s\s'}_{mm'}\bm{\s}^{\s'\s}$ are the Hubbard occupation (or charge) and magnetization, respectively. In Eq.~\eqref{eq:occ_matrix2} and in the following, for the sake of convenience, we use a notation that explicitly highlights the dependence on the magnetization (indicated as a superscript in square brackets), which will be necessary later when we reverse the sign of the magnetization. 
From this representation, using the properties $\trev\,\trev^\dagger = \hat{I}$ (where $\hat{I}$ is the identity operator) and $\trev\,\bm{\s}\,\trev^\dagger=-\bm{\s}$, it follows that:
\begin{equation}
    \begin{split}
        \big(\trev \,\mN_{mm'}^{I[\mM]}\,\trev^\dagger\big)^{\s\s'}& = \sum_{\s_1\s_2}\trev^{\s\s_1} \big(\mN_{mm'}^{I[\mM]}\big)^{\s_1\s_2}\,\trev^{\dagger\s_2\s'}\\
        &=\big(\mN_{m'm}^{I[-\mM]}\big)^{\s\s'},
        \label{chang_occ}
    \end{split}
\end{equation}
which is the transpose to Eq.~\eqref{eq:occ_matrix2} but with the opposite sign in front of the magnetization matrix $\bm{m}^{I}_{mm'}$.
From this expression we can derive how the noncollinear Hubbard potential transforms under the time-reversal operation. Let us start from its definition~\cite{binci_noncollinear_2023}: 
\begin{equation}
\vop_{U}^{[\mM]} = \sum_{Imm'} \frac{U^I}{2} \Big[\delta_{mm'} - 2\mN_{mm'}^{I[\mM]}\Big]\big|\Phi^{I}_{m}\big\rangle\big\langle\Phi^{I}_{m'}\big|,
\label{lin::hubb::pot0}
\end{equation}
where the notation $\vop_{U}^{[\mM]}$ has to be understood as a $2 \times 2$ matrix operator in the spin space. Next, using Eq.~\eqref{chang_occ}, by inverting the dummy indices $m$ and $m'$, and since the application of $\trev$ on $\Phi$ is immaterial (due to the spin-averaging procedure~\cite{binci_noncollinear_2023}) the desired transformation is:  
\begin{equation}
    \begin{split}
    \big(\trev \,\vop_{U}^{[\mM]}\,\trev^{\dagger}\big)^{\s\s'}&= \sum_{\s_1\s_2}\trev^{\s\s_1} \,\big(\vop_{U}^{[\mM]}\big)^{\s_1\s_2}\,\trev^{\dagger\s_2\s'}\\
    &=\big(\vop_{U}^{[-\mM]}\big)^{\s\s'},
    \label{transf_Hubb}
    \end{split}
\end{equation}
which is identical to Eq.~\eqref{lin::hubb::pot0} but with $N_{mm'}^{I[\mM]}$ being replaced by $N_{mm'}^{I[-\mM]}$, which has been discussed above.

\subsubsection*{Dynamical linear response}
 
Let us consider now an external weak perturbation of the system due to the magnetic dynamical potential with a finite $\bq$ and $\omega$ modulation. This potential is given by the interaction energy of the system of electrons with an external magnetic field~\cite{gorni_spin_2018}: 
$V^{[\bm{B}_{\omega\bq}]}_{\mathrm{ext}}(\bq, \omega) = -\mu_\mathrm{B}\,\bm{\sigma}\cdot \bm{B}_{\omega\bq}$.
Hereafter, we indicate with $B_{\omega\bq}^{\alpha}$ the Cartesian $\alpha$-component of the vector amplitude of the external magnetic field. Next, we use the Bloch sum expression for the Hubbard atomic-like states~\cite{cococcioni_accurate_2010, timrov_hubbard_2018}:
\begin{equation}
    |\Phi^s_{m\bm{k}}\rangle=\frac{1}{\sqrt{N_{\bk}}}\sum_l e^{\iota\bm{k}\cdot\bm{R}_l}|\Phi^{ls}_m\rangle=\frac{e^{\iota\bk\cdot\bm{r}}}{\sqrt{N_{\bk}}}|\mF_{m\bm{k}}^s\rangle,
\end{equation}
where $\mF_{m\bm{k}}^s(\bm{r}+\bm{R}_l)=\mF_{m\bm{k}}^s(\bm{r})\equiv\sum_\s \nu_{m\bk}^{s\s}(\bm{r})|\s\rangle$ is the lattice-periodic spinorial part, and $N_{\bk}$ is the number of points in the $\bm{k}$-grid. Here, we used the notation $I\equiv ls$ ($\bm{R}_I \equiv \bm{R}_l+\bm{\tau}_s$), so that $s$ identifies the atomic position within the $l$th cell. Going over to the $(\omega, \bq)$ space and differentiating with respect to the Cartesian $\alpha$-component of $\bm{B}_{\omega\bq}$, we get the linearized Hubbard potential:
\begin{equation}
     \der{\vop_{U,\bk}^{[\mM]}}{\alpha} = -\sum_{smm'} U^s  \der{\mN_{mm'}^{s[\mM]}}{\alpha} \,\big|\mF^{s}_{m\bk+\bq}\big\rangle\big\langle\mF^{s}_{m'\bk}\big|,
    \label{lin::hubb::pot}
\end{equation}
for which a transformation law similar to Eq.~\eqref{transf_Hubb} applies:
\begin{equation}
    \trev \, \derm{\vop_{U,-\bk}^{[\mM]} }{\alpha} \,\trev^\dagger= \der{\vop_{U,\bk}^{[-\mM]}}{\alpha}  .
\end{equation}
We now exploit Bloch's theorem for the KS spinors: $|\Psi_i\rangle\equiv|\Psi_{n\bk}\rangle = \frac{e^{\iota\bk\cdot\bm{r}}}{\sqrt{N}_{\bk}}|\mU_{n\bk}\rangle$. In terms of the lattice-periodic spinorial part $\mU_{n\bk}(\bm{r}+\bm{R}_l)=\mU_{n\bk}(\bm{r})\equiv\sum_\s u^\s_{n\bk}(\bm{r})|\s\rangle$, the first-order response Hubbard occupation matrix in Eq.~\eqref{lin::hubb::pot} is written in terms of a first-order standard and time-reversed response KS wavefunctions:
\begin{equation}
    \begin{split}
        & \der{\mN_{mm'}^{s[\mM]}}{\alpha} = \frac{1}{N_{\bk}} \sum_{n\bk}  \bigg[  \big\langle \mU_{n\bk}\big|\mF_{m'\bk }^s\big\rangle\big\langle\mF_{m\bk+\bq }^s\big|\Delta_{\omega\bq}^\alpha \mU_{n\bk}\big\rangle\\
        & +  \big\langle\trev \mU_{n-\bk}\big|\trev\mF^s_{m-\bk}\rangle\langle\trev\mF^s_{m'-\bk-\bq}\big|\trev\Delta^\alpha_{-\omega-\bq}\mU_{n-\bk}\big\rangle\bigg], \label{lin::hubb::matrix}
    \end{split}
\end{equation}
where the scalar products between the lattice-periodic parts of the spinors are summed over the spin components: $\langle u|\mF\rangle\equiv\sum_\s\langle u^\s|\nu^\s\rangle$. 
Equation~\eqref{lin::hubb::matrix} is valid for finite $\bq$, while for $\bq = \mathbf{0}$ there is an extra term for metallic systems proportional to the derivative of the occupations $\Tilde{\theta}$~\cite{de_gironcoli_lattice_1995, baroni_phonons_2001, timrov_self-consistent_2021}. The implementation of TDDFPT+$U$ does not currently support the case $\bq = \bm{0}$; hence, this term is omitted.
Also, in Eq.~\eqref{lin::hubb::matrix} the prefactor $\tilde{\theta}_{n\bk}$ does not occur due to the definition of the response KS wavefunctions for metallic systems~\cite{de_gironcoli_lattice_1995}.
Thanks to the relations:
\begin{gather*}
  \der{n^{I}_{mm'}}{\alpha} = \mathrm{Tr}\,\bigg[ \der{N^{I}_{mm'}}{\alpha}\bigg], \,\,\,\der{\bm{m}^{I}_{mm'} }{\alpha} = \sum_{\s\s'}  \der{N^{I\s\s'}_{mm'}}{\alpha} \,\bm{\s}^{\s'\s},
\end{gather*}
it is readily seen that the time-reversed part of Eq.~\eqref{lin::hubb::matrix} has an inverted sign for the Hubbard magnetization $\mM$, similarly to what happens for the induced spin-resolved charge density~\cite{urru_density_2019} (in analogy to Eqs.~(26) and (27) of Ref.~\citenum{gorni_spin_2018}). Finally, in Eq.~\eqref{lin::hubb::matrix} the resonant $\Delta_{\omega\bq}^\alpha \mU_{n\bk}$ and the time-reversed antiresonant $\trev\Delta^\alpha_{-\omega-\bq}\mU_{n-\bk}$ wavefunctions can be obtained by solving the two coupled Sternheimer equations, which represent the core methodological development of this work:
%
\begin{strip}
\begin{gather}
        \bigg(\hop_{\bk+\bq}^{[\bB_\mathrm{xc}]} +\vop_{U,\bk+\bq}^{[\mM]} - \epsilon_{n\bk} - \omega \bigg)\Big|\Delta^\alpha_{\omega\bq} \mU_{n\bk} \Big\rangle = -\hat{\mathcal{P}}_{\bk+\bq}\,\Bigg[ \der{\vop_{\mathrm{Hxc}}^{[\bB_\mathrm{xc}]}}{\alpha} + \der{\vop_{U,\bk}^{[\mM]}}{\alpha} + \der{\vop_{\mathrm{ext}}^{ [\bm{B}_{\omega\bq}]}}{\alpha}  \Bigg]\Big|\mU_{n\bk}\Big\rangle, \label{stern::std}\\
        \bigg(\hop_{\bk+\bq}^{[-\bB_\mathrm{xc}]}+\vop_{U,\bk+\bq}^{[-\mM]}-\epsilon_{n-\bk} + \omega\bigg)\Big|\trev \Delta_{-\omega-\bq}^\alpha \mU_{n-\bk} \Big\rangle = -\hat{\Pi}_{\bk+\bq}\Bigg[ \der{\vop_{\mathrm{Hxc}}^{[-\bB_\mathrm{xc}]}}{\alpha} + \der{\vop_{U,\bk}^{[-\mM]}}{\alpha} + \der{\vop_{\mathrm{ext}}^{[-\bm{B}_{\omega\bq}]}}{\alpha}  \Bigg]\Big|\trev \mU_{n-\bk} \Big\rangle, \label{stern::rev}
\end{gather}
\end{strip}
where $\hat{\mathcal{P}}_{\bk+\bq}$ and $\hat{\Pi}_{\bk+\bq}=\trev \hat{\mathcal{P}}_{-\bk-\bq}\trev^\dagger$ are the standard and time-reversed projectors onto the empty electronic states manifold, which have a more complex expression for metallic systems~\cite{de_gironcoli_lattice_1995, baroni_phonons_2001, gorni_spin_2018}. 
Here, $\hop_{\bk+\bq}^{[\bB_\mathrm{xc}]}$ is the ground-state Hamiltonian defined in Eq.~(21) in Ref.~\citenum{gorni_spin_2018}, and $\hop_{\bk+\bq}^{[-\bB_\mathrm{xc}]}$ is its analogue with the reversed sign of $\bB_\mathrm{xc}$, while $\epsilon_{n\bk}$ and $\epsilon_{n-\bk}$ are the ground-state KS energies. This Hamiltonian does not include the ground-state lattice-periodic Hubbard potential $\vop_{U,\bk+\bq}^{[\mM]}$, which instead appears as a separate term in the equations above [see its definition in Eq.~(A15) in Ref.~\citenum{timrov_hubbard_2018}], and $\vop_{U,\bk+\bq}^{[-\mM]}$ is its analogue with the reversed sign for the Hubbard magnetization $\mM$. Finally, the response Hartree and xc (Hxc), response Hubbard, and external potentials appear on the right-hand side of the equations above with the specified signs of $\bB_\mathrm{xc}$, $\mM$, and $\bm{B}_{\omega\bq}$. Equations~\eqref{stern::std} and \eqref{stern::rev} differ from analogues equations in Ref.~\citenum{gorni_spin_2018} by the presence of the ground-state and response Hubbard potentials tuned by the magnitude of the Hubbard $U$ parameter. In addition, the ground-state KS energies, wavefunctions, and spin-resolved charge density are obtained from the DFT+$U$ ground state. Therefore, by solving Eqs.~\eqref{stern::std} and \eqref{stern::rev} self-consistently it is possible to obtain magnon energies including the Hubbard corrections with full internal consistency. 

In principle, the dynamical Sternheimer equations~\eqref{stern::std} and \eqref{stern::rev} can be solved at each frequency $\omega$; this provides the response KS wavefunctions $\Delta^\alpha_{\omega\bq}\mU_{n\bk}$ and the corresponding time-reversed ones $\trev\Delta^\alpha_{-\omega-\bq}\mU_{n-\bk}$. The evaluation of these quantities gives access to the dynamical response spin-charge density matrix operator, defined as: 
\begin{equation}
\begin{split}
    \der{\hat{\rho}}{\alpha} = \frac{1}{N_{\bk}} \sum_{n\bk} &\Big(  \big|\Delta^\alpha_{\omega\bq}\mU_{n\bk}\big\rangle\langle \mU_{n\bk}|\\
&+  |\trev\Delta^\alpha_{-\omega-\bq}\mU_{n-\bk}\big\rangle\langle\trev \mU_{n-\bk}|\Big) .
\label{eq:1densityoperator}
\end{split}
\end{equation}
The knowledge of this latter gives a complete description of the magnetic linear response of the system to a magnetic external perturbation, inasmuch as it delivers the spin susceptibility tensor through: 
\begin{equation}
    \chi_{\alpha\alpha'}(\bq,\omega)=  \mu_\mathrm{B}\, \mathrm{Tr}\,\Bigg[\sigma_\alpha \, \der{\hat{\rho}}{\alpha'} \Bigg].
    \label{eq:chi}
\end{equation}
%
%
%
This dynamical Sternheimer approach was successfully employed in Refs.~\citenum{savrasov_linear_1998, cao_ab_2018} (without Hubbard corrections) to calculate magnons in elemental itinerant metallic magnets and in the context of lattice-dynamical properties~\cite{calandra_adiabatic_2010, giustino_gw_2010, binci_first-principles_2021}. However, the main disadvantage of this approach is its high computational cost coming from the need to solve these equations self-consistently for each value of the frequency $\omega$. To avoid this drawback, we employ the LL approach~\cite{Baroni:2012, gorni_spin_2018}, that enables the determination of the target column of the spin susceptibility tensor along the $\omega$ axis at once with a single linear-response calculation. The technical details concerning the implementation of noncollinear Hubbard functionals within this methodology can be found in the Methods section.

\subsection*{Applications}
In this section, we present the application of the TDDFPT+$U$ implementation using the LL approach to the transition-metal monoxides NiO and MnO. We first discuss the structural and electronic properties of these materials using LSDA and LSDA+$U$, comparing them with experimental data. Next, we present the calculated magnon dispersions using TDDFPT (within ALSDA) and TDDFPT+$U$ (within ALSDA+$U$), and compare these results with experimental estimates. Finally, we extract the Heisenberg exchange interaction parameters by fitting the calculated magnon spectrum and compare them with experimental values. Both NiO and MnO are widely investigated antiferromagnetic (AFM) type II insulators. They crystallize in a rocksalt-type structure in the paramagnetic phase. Below their respective N\'eel temperatures, 524~K for NiO~\cite{Srinivasan:1984} and 120~K for MnO~\cite{Blech:1964}, these materials exhibit a rhombohedral distortion along the [111] direction of the face-centered cubic (fcc) lattice. 

\begin{figure}[]
\centering
\includegraphics[width=0.48\textwidth]{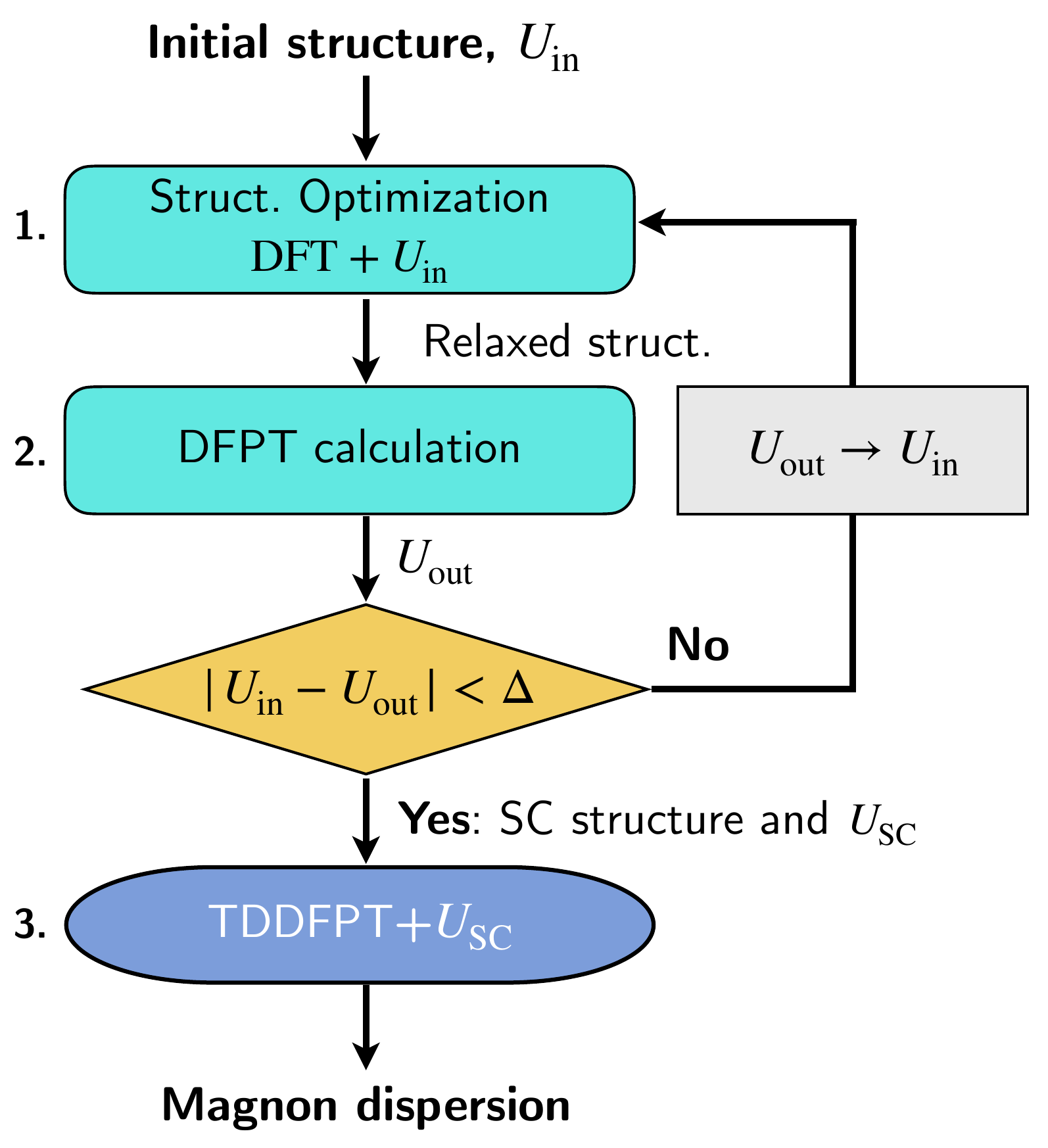}
\caption{\textbf{Computational protocol for the determination of the self-consistently (SC) optimized Hubbard $U$ parameter, crystal structure, and magnon dispersion}. $U_\mathrm{in}$, $U_\mathrm{out}$, and $U_\mathrm{SC}$ are the input, output, and self-consistent Hubbard parameters, respectively, while $\Delta$ is the convergence threshold.}
\label{flowchart}
\end{figure} 

\begin{table}[t]
    \setlength\tabcolsep{0.03in}
    \centering
    \begin{tabular}{clllll}
        \toprule
                     & Method    & $a$ (\AA) & $\vartheta$ (deg) & $|\mM|$ ($\mu_\mathrm{B}$) & $E_g$ (eV) \\
        \midrule
        \multirow{3}{*}{NiO} & LSDA      & 4.93                 & 33.97              & 1.15                   & 0.48     \\
                             & LSDA+$U$  & 5.03                 & 33.63              & 1.60                   & 3.04     \\
                             & Expt.     & 5.11$^\mathrm{a}$    & 33.56$^\mathrm{a}$ & 1.77$^\mathrm{c}$      & 4.0$^\mathrm{e}$  \\
                             &           &                      &                    & 1.90$^\mathrm{d}$               & 4.3$^\mathrm{f}$  \\
        \midrule
        \multirow{3}{*}{MnO} & LSDA      & 5.07                 & 35.66              & 3.86                   & 0.65            \\
                             & LSDA+$U$  & 5.32                 & 34.16              & 4.19                   & 1.93            \\
                             & Expt.     & 5.44$^\mathrm{b}$    & 33.56$^\mathrm{b}$ & 4.79$^\mathrm{c}$      & 4.1$^\mathrm{e}$         \\
                             &           &                      &                    & 4.58$^\mathrm{d}$      & 3.9$\pm$0.4$^\mathrm{g}$ \\
        \bottomrule
    \end{tabular} 
    \caption{\textbf{Crystal and electronic structure properties of NiO and MnO as computed using LSDA, LSDA+$U$, and as measured in experiments}. The equilibrium rhombohedral lattice parameter ($a$), rhombohedral angle ($\vartheta$), magnetic moment ($|\mM|$), and band gap ($E_g$) are presented. The experimental values for $a$ and $\vartheta$ are determined from the cubic lattice using the experimental lattice parameter (4.17 and 4.43~\AA\ for NiO$^\mathrm{a}$ and MnO,$^\mathrm{b}$ respectively), since experimentally the rhombohedral distortion is not quantified. The angle $\vartheta = 33.56^\circ$ corresponds to the case with no rhombohedral distortion. Ref.$^\mathrm{a}$:~\citenum{Schmahl:1964}, Ref.$^\mathrm{b}$:~\citenum{Sasaki:1980}, Ref.$^\mathrm{c}$:~\citenum{Fender:1968}, Ref.$^\mathrm{d}$:~\citenum{Cheetham:1983}, Ref.$^\mathrm{e}$:~\citenum{Kurmaev:2008}, Ref.$^\mathrm{f}$:~\citenum{Sawatzky:1984}, Ref.$^\mathrm{g}$:~\citenum{vanElp:1991}.}
    \label{tab:ground_state}
\end{table}

We start our analysis by determining the electronic and crystal structure of the ground state of NiO and MnO. The $U$ parameter employed for the Hubbard augmentation is calculated using linear-response theory~\cite{cococcioni_linear_2005}:
\begin{equation}
    U^I=\Bigg(\bigg[\frac{dn_0^{I}}{d\lambda_{I'}}\bigg]^{-1}-\bigg[\frac{dn^{I}}{d\lambda_{I'}}\bigg]^{-1}\Bigg)_{II},
    \label{eq:U_from_LRT}
\end{equation}
where $n^I\equiv\sum_{m}\mathrm{Tr}\,(N^{I}_{mm})$ and $n_0^I\equiv\sum_{m}\mathrm{Tr}\,(N^{I}_{0,mm})$ are the interacting and noninteracting response atomic occupations, respectively, which are decomposed into monochromatic components according to DFPT~\cite{timrov_hubbard_2018,binci_noncollinear_2023}: ${dn^{I}}/{d\lambda_{I'}}=\frac{1}{N_{\bq}}\sum_{\bq}^{N_{\bq}}e^{\iota\bq\cdot(\bm{R}_l-\bm{R}_{l'})}\Delta_{\bq}^{s'}n^s$  
(detailed expression can be found in Refs.~\citenum{timrov_hubbard_2018, timrov_self-consistent_2021, binci_noncollinear_2023}).
In Eq.~\eqref{eq:U_from_LRT}, the Hubbard $U$ parameter is defined by requiring that the second derivative of the DFT-LSDA total energy with respect to the occupation of the Hubbard manifold is zero. This condition eliminates the unphysical curvature of the total energy caused by SIEs, thereby ensuring the piecewise linearity of the LSDA+$U$ total energy~\cite{cococcioni_linear_2005}.
Thanks to the collinearity of the AFM ground state, and the neglect of the spin-orbit coupling due to the lightness of the elements, it is safely possible to restrict such a calculation to the collinear case, thus saving substantial computational effort. In order to simultaneously optimize the Hubbard parameters and the crystal structure, we employed the workflow proposed in Refs.~\citenum{cococcioni_energetics_2019,timrov_self-consistent_2021} and depicted in Fig.
1. This latter alternates variable-cell structural relaxations and determination of the $U$ parameter until convergence is achieved. We obtained self-consistent $U$ values of 6.26 and 4.29~eV for Ni-$3d$ and Mn-$3d$ states in NiO and MnO, respectively. Finally, these values are used to compute the magnon dispersions using TDDFPT+$U$.

Table~\ref{tab:ground_state} summarizes the equilibrium rhombohedral lattice parameter ($a$), rhombohedral angle ($\vartheta$), magnetic moment ($|\mM|$), and band gap ($E_g$) as computed using LSDA, LSDA+$U$, and as measured in experiments (see also Fig. 4b
). The experimental values for $a$ and $\vartheta$ are derived from the cubic lattice~\cite{Schmahl:1964, Sasaki:1980}. We are not aware of any direct experimental reports of the rhombohedral lattice parameters for these materials, possibly because the rhombohedral distortions are small and hard to resolve experimentally. Therefore, the comparison of our theoretical rhombohedral lattice parameters with the experimental reference values should be considered somewhat peripheral.  
We find that the Hubbard correction systematically improves over LSDA. Still, non-negligible discrepancies are present even in the $\mathrm{LSDA}+U$ approach, which are to a large extent due to the limitation of the base xc LSDA functional that is known to cause excessive binding in crystal structures. In Sec.~S1 of the supplemental information (SI), we show how the structural properties of the two systems change when the base xc functional is replaced with spin-polarized PBE~\cite{Perdew:1996} or PBEsol~\cite{Perdew:2008}, combined with their respective self-consistent $U$ values computed via DFPT. We find that PBE+$U$ slightly overestimates the lattice parameter $a$, while PBEsol+$U$ slightly underestimates it, with only minor variations in $\alpha$. Overall, PBEsol+$U$ provides more accurate predictions of the crystal structure than LSDA+$U$. Since our TDDFPT+$U$ implementation currently does not support $\sigma$-GGA functionals, the remainder of this study focuses exclusively on LSDA+$U$. 
Concerning $E_g$, although its value generally improves when Hubbard corrections are included (see e.g. Ref.~\citenum{KirchnerHall:2021}), an underestimation of this quantity is to be expected, inasmuch as $\mathrm{DFT}+U$ is a theoretical framework that mainly corrects total energies. More precisely, when the band edges have the same orbital character of the Hubbard projectors, reasonable band gaps can be obtained with Hubbard-corrected functionals~\cite{Lambert:2023, leiria_campo_jr_extended_2010, Lee:2020}. However, in the general case, for more accurate evaluations of spectral properties, methods like $GW$\cite{vanSetten2015, Grumet2018, Bonacci2023}, hybrid functionals\cite{wing_band_2021,Skone2016, Chen2018}
or Koopmans functionals~\cite{Dabo:2010, Nguyen:2018, Linscott:2023} may be more appropriate.

\begin{figure}[t]
\centering
\includegraphics[width=0.45\textwidth]{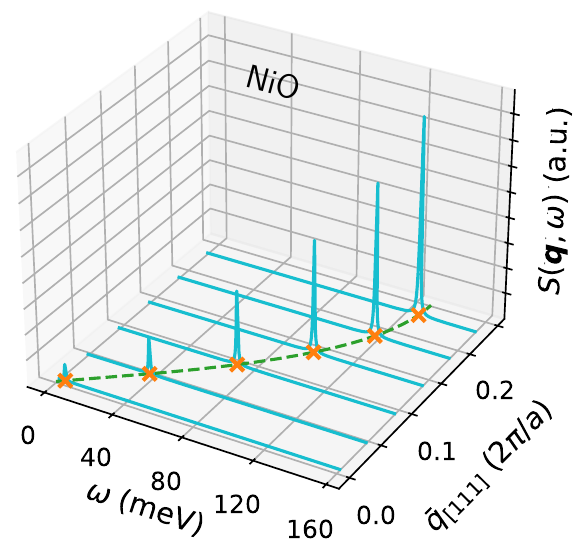}
\caption{\textbf{Magnetic spectrum of NiO}. Calculated dynamical spin structure factor from Eq.~\eqref{eq:Sq} (in cyan color) as a function of the frequency $\omega$ (in meV) at several values of the transferred momentum $q$ along the [111] direction (in units of $2\pi/a$, where $a$ is the rhombohedral lattice parameter obtained from the distorted cell within LSDA+$U$). Orange crosses mark the exact position of magnon peaks, and the green dashed line is a guide for the eye to highlight the magnon dispersion in the $(\omega,q)$ plane.}
\label{fig:S_q}
\end{figure} 

Going over to the calculation of magnons, we recall that in the first Born approximation, there is a relation between the experimentally detectable double-differential cross section $d^2\sigma/(d\Omega d\omega)$ (measuring the scattering of neutrons) and the spin susceptibility tensor $\bm{\chi}(\bq,\omega)$ which is given by: $d^2\sigma/(d\Omega d\omega)=-\frac{g_\mathrm{n}^2}{4\pi}\frac{k_\mathrm{f}}{k_\mathrm{i}}\, \mathcal{S}(\bq,\omega)$ \cite{van_hove_correlations_1954,blume_polarization_1963,gorni_spin_2018}, where $g_\mathrm{n}$ is the neutron $g$-factor, $k_\mathrm{i}$ and $k_\mathrm{f}$ are the initial and final wavevectors of the scattered neutrons, and
\begin{equation}
   \mathcal{S}(\bq,\omega)=  -\mathrm{Im}\,\mathrm{Tr}\,\Big(\bm{T}(\bq)\bm{\chi}(\bq,\omega)\Big),
   \label{eq:Sq}
\end{equation}
where $T_{\alpha\alpha'}(\bq)=\delta_{\alpha\alpha'}-q_\alpha q_{\alpha'}/q^2$ is the projector to a plane transverse to the transferred momentum $\bq$, and $q=|\bq|$. The poles of $\mathcal{S}(\bq,\omega)$ occur at the frequencies $\omega_\mathrm{m}(\bq)$ of magnons and Stoner excitations. To illustrate how the determination of the magnon spectrum is carried out, we report in Fig. 
2 an example of the calculated $\mathcal{S}(\bq,\omega)$ for different values of the transferred momenta $\bq$ for the magnon branch of NiO along the [111] direction. The sharp resonances in $\mathcal{S}(\bq,\omega)$ indicate the frequency positions of the magnetic excitation of the system, which draw the spin-wave dispersions in the $(\omega,\bq)$ plane.

\begin{figure}[t]
\centering
\includegraphics[width=0.5\textwidth]{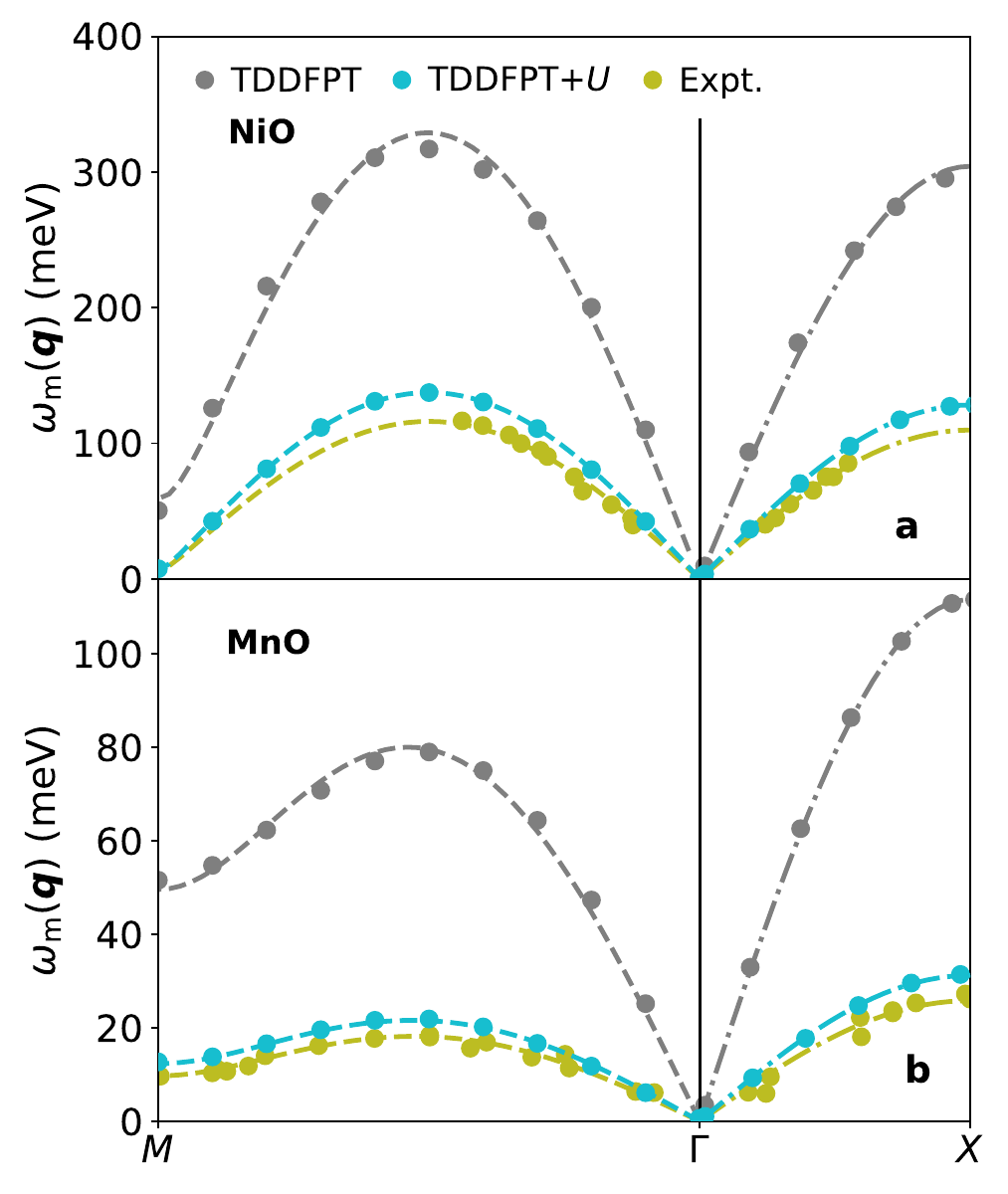}
\caption{\textbf{Calculated magnon dispersions with different methods and comparison with experiments}. (a)~NiO and (b)~MnO computed using TDDFPT (grey dots), TDDFPT+$U$ (cyan dots), and as measured in experiments (olive dots)~\cite{hutchings_measurement_1972, pepy_spin_1974}. Dashed lines along $\Gamma - M$ and $\Gamma - X$ are fit using Eq.~\eqref{eq:omega_fit} with the parameters $J^+_1$, $J^-_1$, and $J_2$. The values of the magnetic exchange parameters resulting from the fit are summarized in Table~\ref{tab:exchange_param}}
\label{fig:magnon_spectrum}
\end{figure} 
\begin{figure}[t]
\centering
\includegraphics[width=0.48\textwidth]{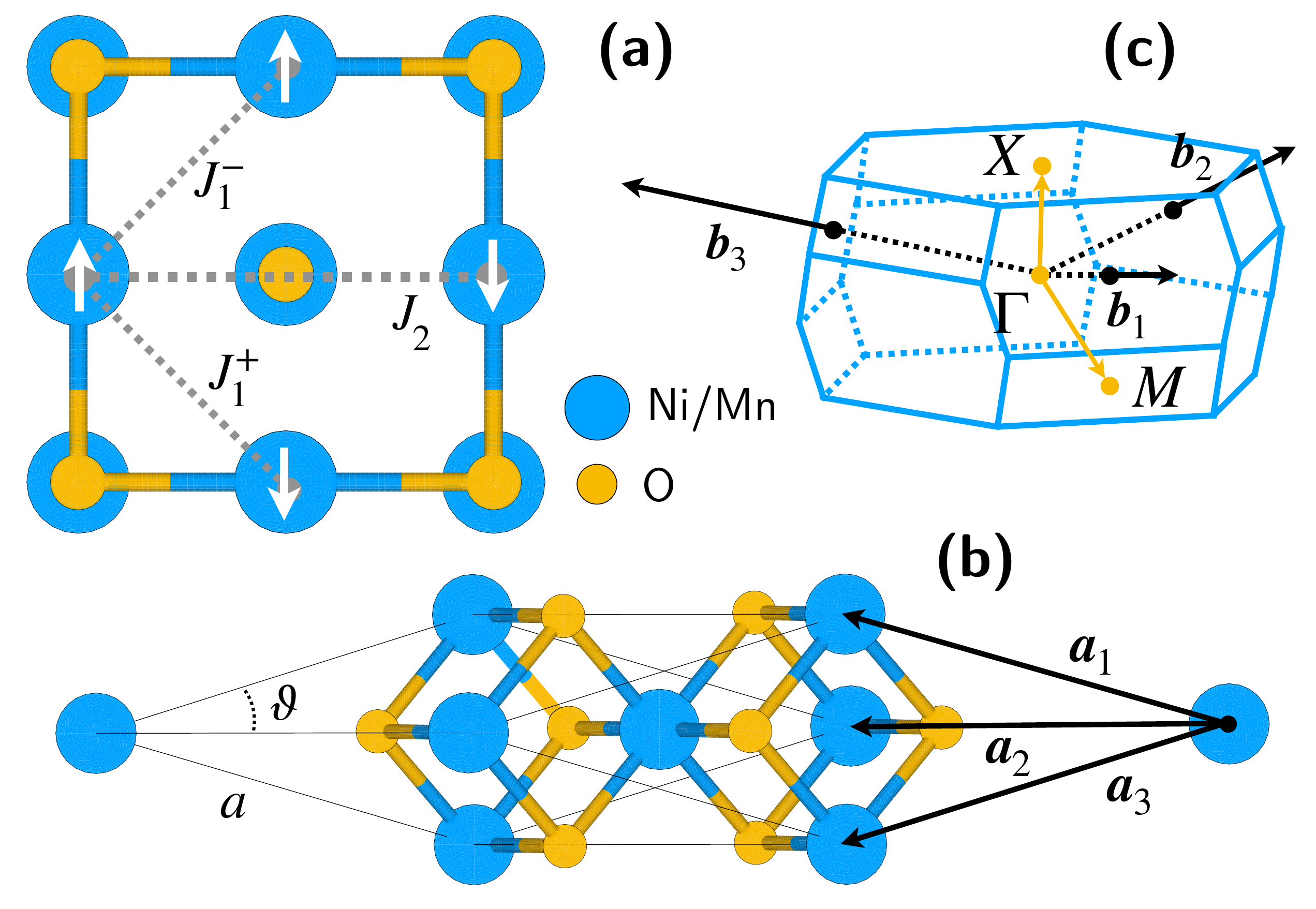}
\caption{\textbf{Schematic illustration of the NiO and MnO unit cell}. In the picture, the exchange interaction parameters $J^+_1$, $J^-_1$, and $J_2$ introduced in Eq. (\ref{eq:heisenberg_Hamiltonian}), the lattice parameter $a$, and the angle $\vartheta$ quantifying the rhombohedral distortion are reported. Panel (a) shows the fcc cell, while panel (b) displays the rhombohedral cell employed in our first-principles calculations, and (c) is the Brillouin zone (BZ) corresponding to the rhombohedral cell. $\bm{a}_1$, $\bm{a}_2$, and $\bm{a}_3$ are the real-space primitive lattice vectors of the rhombohedral unit cell, while $\bm{b}_1$, $\bm{b}_2$, and $\bm{b}_3$ are the reciprocal-space primitive lattice vectors of the BZ. The high-symmetry points $\Gamma$, $X$, and $M$ in the BZ are also highlighted.}
\label{fig:sketch_magn}
\end{figure} 

In Fig. 3
we show the calculated magnon spectra for NiO and MnO with and without the Hubbard correction, and the comparison with experimental measurements \cite{hutchings_measurement_1972,pepy_spin_1974}. Plain TDDFPT using ALSDA significantly overestimates the magnon energies with respect to experiments for both NiO and MnO. This aligns with previous LSDA-based theoretical studies~\cite{wan_calculation_2006, kotani_spin_2008}. On the contrary, the effect of the Hubbard augmentation  is substantial, and ALSDA+$U$ significantly improves agreement with experiments, not only for the amplitude of the magnon dispersion, but also for their curvature. This outcome is remarkable since the TDDFPT+$U$ calculations are fully first-principles and do not rely on any empirical parametrization for the lattice parameters, $U$ value, or type and/or strength of magnetic interactions. For MnO, we remark that the inclusion of the rhombohedral distortion is crucial in order to obtain a nonzero magnon energy at the $M$ point~\cite{jacobsson_exchange_2013,fischer_exchange_2009}. For NiO instead the rhombohedral distortion is much smaller, resulting in an almost vanishing magnon energy at $\bq = M$ as compared to the magnon bandwidth. We observe that it is important to account for the rhombohedral distortion within the selection of the $\bq$--path across the Brillouin zone (BZ): the $\Gamma - M$ high-symmetry direction undergoes a small rotation, while only the length along the $\Gamma - X$ line is slightly modified. To do this, we specified the coordinates of the $X$ and $M$ points in the crystal framework (i.e., in the basis of the reciprocal lattice vectors of the BZ) and then transformed them to the Cartesian framework.

The magnon energy at $\bq \rightarrow \mathbf{0}$ requires special attention due to the Goldstone theorem, which dictates that the acoustic magnon energy must vanish in the absence of spin-orbit coupling~\cite{Goldstone:1962, Watanabe:2012}. However, in practice, numerical calculations often violate this condition due to approximations used to describe the ground and excited states, such as differing $\bk$-point grids or basis sets~\cite{rousseau_efficient_2012, Lounis:2010, Lounis:2011, Buczek:2011b, skovhus_dynamic_2021, Muller:2016}. In contrast, our TDDFPT+$U$ implementation based on the LL approach satisfies the Goldstone theorem with high accuracy and correctly reproduces the long-wavelength limit, as previously demonstrated for standard TDDFPT with ALSDA~\cite{gorni_turbomagnon_2022}. Since our current implementation does not support the exact $\bq = \mathbf{0}$ limit, we perform calculations using very small but finite $\bq$ values near $\mathbf{0}$. As shown in Sec.~S3 of the SI, we find that the magnon energy indeed vanishes as $\bq \rightarrow \mathbf{0}$ in the absence of spin-orbit coupling. In contrast, the TDDFT+$U$ implementation in Refs.~\citenum{Skovhus:2022, Skovhus:2022b}, which solves the Dyson equation, employs the so-called $\lambda$ALSDA+$U$, where $\lambda$ is an empirical rescaling factor applied to the xc kernel to enforce the Goldstone condition. For antiferromagnetic insulators such as Cr$_2$O$_3$, a $\lambda$ value of 1.4 was required to achieve a zero magnon energy at $\bq = \mathbf{0}$. However, when using empirical $U$ values greater than 1~eV, unphysical behavior was observed in the acoustic magnon branch, including vanishing energy over extended portions of high-symmetry directions in the BZ~\cite{Skovhus:2022}. This demonstrates the challenge of simultaneously satisfying the Goldstone condition and achieving magnon dispersions in close agreement with experimental results when using $\lambda$ALSDA+$U$. The approach in Refs.~\citenum{Lounis:2010, Lounis:2011}, which enforces sum rules to satisfy the Goldstone condition, could potentially complement the TDDFT+$U$ Dyson-based method of Refs.~\citenum{Skovhus:2022, Skovhus:2022b} to address this issue. However, such corrections are unnecessary in our TDDFPT+$U$ LL-based approach, which satisfies the Goldstone condition because the same computational parameters enter in both the DFT+$U$ ground state and in the linear-response calculations. This highlights the robustness and accuracy of our novel method.

To gain further physical insights, we fit the magnon dispersions of Fig. 
3 in order to have access to the Heisenberg exchange interaction parameters. For this analysis, we consider a Heisenberg model with nearest-neighbour (n.n.) and next-nearest-neighbour (n.n.n.) exchange interactions. The low-energy Hamiltonian reads~\cite{pepy_spin_1974}:
\begin{equation}
    \begin{split}    
    \hat{H} = &\sum_{i,j}^\mathrm{n.n.p} J_1^{-}\,{{\bm{S}}}_i\cdot{\bm{S}}_{j} + \sum^\mathrm{n.n.a}_{i,j}J_1^{+}\,{\bm{S}}_i\cdot{\bm{S}}_{j} + \sum_{i,j}^\mathrm{n.n.n.}J_2\,{\bm{S}}_i\cdot{\bm{S}}_{j},
    \label{eq:heisenberg_Hamiltonian}
    \end{split}
\end{equation}
where the first sum is over nearest-neighbors with parallel spins ($J_1^{-}$), the second sum is over nearest-neighbors with antiparallel spins ($J_1^{+}$), and the third sum is over next-nearest-neighbors with antiparallel spins ($J_2$) (see Fig. 
4a). The notation $J_1^{-}$ and $J_1^{+}$, introduced by Lines and Jones~\cite{lines_antiferromagnetism_1965}, highlights the difference in interaction strength for nearest neighbors due to rhombohedral distortions~\cite{Kohgi:1972}. Without these distortions, $J_1^{-} = J_1^{+}$. The Heisenberg Hamiltonian in Eq.~\eqref{eq:heisenberg_Hamiltonian} uses the same convention as in Ref.~\citenum{pepy_spin_1974}, assuming that more remote exchange interactions are negligible. We also neglect DM and single-ion anisotropy magnetic interactions in this Hamiltonian. Here, $i$ and $j$ label magnetic atomic sites, and $\sum_{i,j}$ denotes the summation over pairs $(i,j)$.

\begin{table*}[t!]
    \renewcommand{\arraystretch}{1.}
    \setlength\tabcolsep{0.1in}
    \centering
    \begin{tabular}{clccccccc}
        \toprule
        Material             & Method        & $J_1^+$  & $J_1^-$  &  $J_1$  & $J_2$   & $|J_2/J_1|$ & $\Delta J_1$ & $|\Delta J_1/J_1| \times 100\%$ \\
        \midrule
        \multirow{3}{*}{NiO} & TDDFPT        & $-$2.89  & $-$3.20  & $-$3.05 & 28.25   &    9.26     & 0.31         & 10.2\%             \\
                             & TDDFPT+$U$    & $-$1.18  & $-$1.19  & $-$1.19 & 11.87   &    9.97     & 0.01         &  0.8\%             \\
                             & Expt.         & $-$0.81  & $-$0.82  & $-$0.82 &  9.96   &   12.14     & 0.01         &  1.2\%             \\
        \midrule
        \multirow{3}{*}{MnO} & TDDFPT        &  1.81    & 1.21     &  1.51   & 1.91   &    1.26     &  0.60        &  39.7\%            \\
                             & TDDFPT+$U$    &  0.51    & 0.37     &  0.44   & 0.53   &    1.20     &  0.14        &  31.8\%            \\
                             & Expt.         &  0.40    & 0.30     &  0.35   & 0.46   &    1.31     &  0.10        &  28.6\%            \\
        \bottomrule
    \end{tabular}
    \caption{\textbf{Exchange interaction parameters}. Exchange interaction parameters (in meV), using the convention for the Heisenberg Hamiltonian of Eq.~\eqref{eq:heisenberg_Hamiltonian}, are extracted by fitting the magnon dispersions of Fig.
    3 using TDDFPT, TDDFPT$+U$, and experimental data from Refs.~\citenum{hutchings_measurement_1972} and \citenum{pepy_spin_1974} for NiO (at 78~K) and MnO (at 4~K), respectively. Here, $J_1 = (J_1^+ + J_1^-)/2$, and $\Delta J_1 = J_1^+ - J_1^-$.}
    \label{tab:exchange_param}
\end{table*}

From the Heisenberg Hamiltonian Eq.~\eqref{eq:heisenberg_Hamiltonian}, using LSWT, it is possible to obtain an analytical expression for the magnon dispersion $\omega_\mathrm{m}(\bq)$ that explicitly depends on the exchange interaction parameters, which can then be used to fit the magnon spectrum (Fig. 3
). The initial step involves transforming the Hamiltonian to represent each spin within its local reference frame, oriented along the $z$-direction. Next, a Holstein-Primakoff transformation is performed~\cite{Holstein:1940}, replacing spin operators with creation and annihilation bosonic operators similar to those in a harmonic oscillator. During this transformation, the Hamiltonian is linearized, retaining terms up to the second order in the Holstein-Primakoff bosonic quasiparticles. Finally, the equation of motion is solved by diagonalizing the dynamical matrix, and a Bogoliubov transformation ensures a diagonalizing basis that adheres to the usual commutation relations. The final stage delivers the analytical expression for the magnon frequencies, $\omega_\mathrm{m}(\bq)$. For a fcc lattice, this expression reads~\cite{lines_antiferromagnetism_1965}:
\begin{equation}
    \begin{split}
        \omega_\mathrm{m}(\boldsymbol{q})&=\mu\,\sqrt{\big[J_{11}(\bm{q})-J_{11}(\bm{0})+J_{12}(\bm{0})\big]^2-J_{12}(\bm{q})^2},
        \label{eq:omega_fit}
    \end{split}
\end{equation}
where $\mu = 2 S$, i.e. it depends on the nominal magnetic moment $S$ of the transition-metal ion ($S=1$ for Ni$^{2+}$ and $S=5/2$ for Mn$^{2+}$). The functions $J_{11}(\bm{q})$ and $J_{12}(\bm{q})$ are defined as~\cite{hutchings_measurement_1972, pepy_spin_1974, Factor2}:
\begin{gather*}
        J_{11}(\bm{q}) = J_1^{-}\,\sum_{\alpha \neq \alpha'} \cos\pi(q_\alpha-q_{\alpha'}), \\
        J_{12}(\bm{q}) = 2 J_2\sum_\alpha \cos\, (2\pi q_\alpha) + J_1^+\,\sum_{\alpha \neq \alpha'} \cos\pi(q_\alpha+q_{\alpha'}).
\end{gather*}
In Fig. 3
, we show the fit of the magnon dispersions of NiO and MnO from TDDFPT and TDDFPT+$U$ using Eq.~\eqref{eq:omega_fit} simultaneously along the $\Gamma - M$ and $\Gamma - X$ directions. Table~\ref{tab:exchange_param} compares the theoretical Heisenberg exchange parameters obtained by fitting the TDDFPT and TDDFPT+$U$ magnon dispersions in Fig. 3
with experimental values~\cite{hutchings_measurement_1972, pepy_spin_1974}. 
We extract the experimental Heisenberg exchange parameters by fitting the experimental magnon dispersions using Eq.~\eqref{eq:omega_fit}, omitting single-ion anisotropy terms and using the $S$ values given after Eq.~\eqref{eq:omega_fit}. Our fitted values differ slightly from those in Refs.~\citenum{hutchings_measurement_1972, pepy_spin_1974} because the original works included single-ion anisotropy and used a slightly different $S$ value for MnO. This refitting ensures a consistent comparison between theoretical and experimental exchange parameters under similar conditions and approximations.
As shown in Table~\ref{tab:exchange_param}, the $J_1^+$, $J_1^-$, and $J_2$ parameters from TDDFPT within ALSDA are significantly overestimated compared to experimental values, while TDDFPT+$U$ provides parameters much closer to experiments. 
Within TDDFPT, the theoretical exchange parameters deviate from the experimental ones by $150-500\%$, while within TDDFPT+$U$ they deviate by $22-33\%$, drastically improving the accuracy of the predictions. It is worth stressing that the sign of these parameters is correct in both cases. Given the strong dependence of the parameters $J$ on the structural properties, we observe once again that the residual disagreement obtained with the Hubbard corrections could be further lowered by using more advanced xc functionals~\cite{perdew_restoring_2008,sun_strongly_2015}. It would be possible to bring the theoretical exchange parameters even closer to experimental values by slightly increasing the $U$ parameter\cite{jacobsson_exchange_2013}, but this would introduce an adjustable parameter and, by doing so, the theory would not be unbiased anymore. This also highlights the strong sensitivity of the magnon dispersions and the corresponding $J$ parameters to the value of the Hubbard $U$.

The difference between $J_1^+$ and $J_1^-$ serves as a useful measure of the impact of rhombohedral distortions on the magnon dispersions. We define $\Delta J_1 = J_1^+ - J_1^-$, and its relative strength compared to the average $J_1$ parameter as $|\Delta J_1/J_1|$. Table \ref{tab:exchange_param} shows that $\Delta J_1$ and $|\Delta J_1/J_1|$ are in much closer agreement with the experimental values when using TDDFPT+$U$ as compared to TDDFPT.
When $\Delta J_1 \rightarrow 0$, the magnon energy at the $M$ point in the BZ vanishes, while increasing values of $\Delta J_1$ lead to higher magnon energy at the $M$ point, which can be verified using Eq.~\eqref{eq:omega_fit}. This implies that $\Delta J_1$ reflects the crystallographic inequivalence between two nearest-neighbor transition-metal ions, which is influenced by the rhombohedral distortion. Our first-principles calculations are in agreement with this picture, predicting a larger rhombohedral angle $\vartheta$ (indicating greater rhombohedral distortion) for MnO compared to NiO (see Table~\ref{tab:ground_state}). Consistently, LSDA, which overestimates $\vartheta$ for both materials, displaying the largest values of both $\Delta J_1$ and the magnon energy at the $M$ point. 

\section*{DISCUSSION}

We have presented a first-principles approach for calculating magnons based on time-dependent density-functional perturbation theory~\cite{gorni_spin_2018} and the Liouville-Lanczos method augmenting the adiabatic exchange-correlation functional with nonempirical Hubbard corrections. This Hubbard-extended formulation of TDDFPT is fully \textit{ab initio}, since the Hubbard $U$ parameter is computed from first-principles using DFPT~\cite{timrov_hubbard_2018, binci_noncollinear_2023}, avoiding any empirical calibrations. Additionally, the dynamical spin susceptibility tensor is directly computed through linear-response theory, without assumptions about the type and strength of magnetic interactions, unlike spin models such as the Heisenberg Hamiltonian. The Hubbard $U$ correction is included self-consistently when solving the ground-state DFT+$U$ problem, and its linear response is included when solving the TDDFPT+$U$ equations using the LL approach. We chose the LL approach over the Dyson or Sternheimer methods because it is computationally efficient and provides access to all frequencies at once, unlike the point-wise calculations required by the other methods. Similarly to the Sternheimer approach~\cite{cao_ab_2018}, the LL approach avoids the computationally expensive summations over empty states~\cite{gorni_spin_2018}, commonly performed in static DFPT for phonons~\cite{baroni_greens-function_1987}. In all cases, the local spin density approximation (adiabatic in the time-dependent density-functional perturbation theory case) is used for the base exchange-correlation functional, then augmented with the Hubbard corrections. Crucially, the Goldstone theorem is satisfied within this computational framework.

To benchmark the TDDFPT+$U$ formalism and ensure the correctness of its implementation, we applied it to the prototypical transition-metal monoxides NiO and MnO, including their rhombohedral lattice distortions. The computed magnon dispersions from TDDFPT+$U$ show remarkable agreement with experimental data, unlike those from TDDFPT. Specifically, we accurately predict finite magnon energy at the $M$ point in the BZ of MnO due to rhombohedral distortions, while this effect is almost negligible for NiO with respect to its magnon bandwidth, consistent with experimental observations. Using the Heisenberg Hamiltonian and the LSWT, we fit the magnon dispersions to extract the nearest-neighbor and next-nearest-neighbor Heisenberg interaction parameters. The parameters obtained from TDDFPT+$U$ align well with experimental values, whereas those from TDDFPT are largely overestimated. A detailed comparison of Heisenberg exchange parameters from TDDFPT+$U$ and those calculated via total energy differences or the infinitesimal-rotations method is of great interest and is presented in a separate study~\cite{dosSantos:2024}.

Although the current implementation of TDDFPT+$U$ using the LL approach yielded good results, it still has limitations. Indeed, this approach is based on the linear-response regime, meaning it cannot simulate ultrafast phenomena with strong external perturbations~\cite{tancogne-dejean_time-dependent_2020}. Moreover, when solving the dynamical TDDFPT+$U$ equations, the response Hubbard potential is computed using the static Hubbard $U$ parameter, neglecting its dynamical variations due to external perturbations. We believe that this is a reasonable approximation since the external perturbation is assumed to be weak. Investigating the effect of dynamical modulation of $U$ on magnons could be interesting, as done in Ref.~\citenum{tancogne-dejean_ultrafast_2018} for studying high-harmonic generation. However, the $U(\omega)$ parameter cannot be treated within the LL approach because it does not allow for a frequency-dependent response potential. To explore $U(\omega)$, switching to the Dyson or Sternheimer approaches would be necessary. On a more technical level, the current implementation is limited to LSDA, with $\sigma$-GGA currently unsupported, and it works only with norm-conserving pseudopotentials. Extensions to ultrasoft pseudopotentials~\cite{vanderbilt_soft_1990} and the PAW method~\cite{blochl_projector_1994} would be straightforward but they increase the computational complexity. In addition, the current implementation does not support symmetry, requiring the use of the full $\bk$ point grid in the BZ. We plan to implement symmetries in future versions of the TDDFPT+$U$ code, which will further reduce the computational cost. Finally, the current implementation runs only on central processing unit (CPU) architectures, and  porting it to graphics processing unit (GPU) architectures would significantly boost the speed of magnon calculations~\cite{Giannozzi:2020, Carnimeo:2023}.

Finally, we discuss the outlook and future prospects. Our noncollinear TDDFPT+$U$ implementation, based on Ref.~\citenum{gorni_spin_2018}, supports spin-orbit coupling (SOC). This is particularly relevant for heavy elements such as e.g. rare earths, containing localized $f$ electrons requiring Hubbard corrections.
Moreover, incorporating SOC enables the study of its effect on magnons, e.g. through the magnon-phonon coupling, intrinsic damping, anisotropy and gaps in the magnon spectrum~\cite{gorni_first-principles_2023, delugas_magnon-phonon_2023}.
Thanks to the noncollinear extension of DFT+$U$ and DFPT~\cite{binci_noncollinear_2023}, it is possible to evaluate the Hubbard $U$ parameter within the noncollinear framework, fully including SOC and further employ it within TDDFPT+$U$. 
Moreover, the current TDDFPT+$U$ implementation can be straighforwardly extended to incorporate inter-site Hubbard $V$ corrections~\cite{leiria_campo_jr_extended_2010}, that have proven to be very accurate and effective for diverse materials and properties~\cite{Timrov:2020c, Mahajan:2021, Mahajan:2022, Timrov:2022b, Timrov:2023, Binci:2023, Gebreyesus:2023, Haddadi:2024}. 
Another promising direction involves using TDDFPT+$U$ in combination with MBPT~\cite{Paischer:2023} to study the renormalization and damping of electronic band structures in antiferromagnetic insulators due to electron-magnon coupling~\cite{Muller:2019, Nabok:2021}, with a particular focus on employing \textit{ab initio} rather than empirical $U$ values~\cite{Paischer:2024}.
Additionally, the current TDDFPT+$U$ implementation can be used for high-throughput calculations of magnons for hundreds or even thousands of materials using platforms like AiiDA~\cite{Huber:2020, Bastonero:2024}. This process can be further streamlined using equivariant neural networks to predict Hubbard parameters essentially at no cost, but with the accuracy close to that of DFPT~\cite{Uhrin:2024}. We believe that the present TDDFPT+$U$ extension opens the door to accurate modeling of magnons in complex transition-metal and rare-earth compounds, potentially leading to significant technological breakthroughs in spintronics and magnonics.

\section*{METHODS}
\subsubsection*{Liouville-Lanczos approach with Hubbard corrections}
The LL approach aims to solve the quantum Liouville spinorial equation, which is equivalent to the coupled dynamical Sternheimer equations~\eqref{stern::std} and \eqref{stern::rev}. This is done by linearizing the Liouville equation, and taking advantage of the \emph{batch representation}~\cite{Baroni:2012, walker_efficient_2006, timrov_electron_2013} to cast the equation in a matrix form that can be efficiently solved using the Lanczos algorithm~\cite{Saad:2003}. The key advantage of using the LL approach over directly solving the dynamical Sternheimer equations is that the problem only needs to be solved once, regardless of frequency. The evaluation of the magnetic spectrum is then an inexpensive post-processing step~\cite{Rocca:2008}. The linearized quantum Liouville spinorial equation in the frequency domain reads~\cite{gorni_spin_2018}:
\begin{equation}
     \der{\hat{\rho}}{\alpha} = \big(\omega - \hat{\mathcal{L}}_{\bq} \big)^{-1}\Bigg[  \der{\vop_{\mathrm{ext}}}{\alpha}, \hat{\rho}_0 \Bigg],
    \label{eq:rho_liouville}
\end{equation}
where $\hat{\rho}_0$ is the ground-state spin-charge density matrix operator, $d\vop_{\mathrm{ext}}/dB_{\omega\bq}^\alpha \rightarrow d\vop^{[\pm\bm{B}_{\omega\bq}]}_{\mathrm{ext}}/dB_{\omega\bq}^\alpha $ are the direct and reversed magnetic perturbations, and $\hat{\mathcal{L}}_{\bq} \rightarrow \hat{\mathcal{L}}_{\bq}^{\pm}$ is the Liouvillian superoperator, which action over a generic quantum-mechanical operator $\hat{X}$ is defined as:
\begin{eqnarray}
\hat{\mathcal{L}}_{\bq}^{\pm} (\hat{X}) & \equiv &  
\big[ \hop^{[\pm \bB_\mathrm{xc}]}, \hat{X}\big] + 
\big[ \vop_{U}^{[\pm \mM]}, \hat{X}\big] \\ \nonumber 
& & + \bigg[ \der{\vop_{\mathrm{Hxc}}^{[\pm \bB_\mathrm{xc}]}}{\alpha}, \hat{\rho}_0\bigg]
+ \bigg[ \der{\vop_{U}^{[\pm \mM]}}{\alpha}, \hat{\rho}_0\bigg].
\label{eq:Liouvillian0}
\end{eqnarray}
The first and third terms in the above equation are the noninteracting and interacting terms found in the standard TDDFPT formulation within ALSDA~\cite{gorni_spin_2018}. The second and fourth terms are new, referred to as the noninteracting and interacting Hubbard terms, which arise from the Hubbard $U$ correction. These four terms also appear in the dynamical resonant and antiresonant Sternheimer equations~\eqref{stern::std} and \eqref{stern::rev} with the respective signs of $\bB_\mathrm{xc}$ and $\mM$.

To evaluate the spin susceptibility tensor using Eq.~\eqref{eq:chi}, we can use the solution of Eq.~\eqref{eq:rho_liouville}. For practical computation using the Lanczos algorithm, it is convenient to rewrite Eqs.~\eqref{eq:chi} and \eqref{eq:rho_liouville} as a resolvent of the Liouvillian~\cite{Rocca:2008}:
\begin{equation}
    \chi_{\alpha\alpha'}(\bq, \omega) = \big\langle\mathsf{w}_{\alpha}\big|\big(\omega -\hat{\mathcal{L}}_{\bq} \big)^{-1}\big|\mathsf{v}_{\alpha'}\big\rangle,
    \label{eq:resolvent}
\end{equation}
where $\langle\mathsf{w}_{\alpha}|=\big\langle\{\mU_{n\bk}\}\big| \mu_\mathrm{B}\sigma_{\alpha}$ and $|\mathsf{v}_{\alpha'}\rangle=\big[ \der{\vop_{\mathrm{ext}}}{\alpha'}, \hat{\rho}_0\big]\big|\{\mU_{n\bk}\}\rangle$, while $\{\mU_{n\bk}\}$ denotes a set of the ground-state KS wavefunctions. 
Next, we introduce the dual basis $\{ \langle\mathsf{p}_n|,|\mathsf{q}_n\rangle \}$ composed of ``left'' $\langle\mathsf{p}_n|$ and ``right'' $|\mathsf{q}_n\rangle$ Lanczos orthonormal vectors ($\langle\mathsf{p}_n|\mathsf{q}_{m}\rangle=\delta_{n,m}$) that tridiagonalizes the Liouvillian: $\big\langle\mathsf{p}_n\big| \hat{\mathcal{L}}_{\bq} \big|\mathsf{q}_m\big\rangle = \alpha_n\,\delta_{n,m}+\beta_n\,\delta_{n,m+1}+\gamma_n\,\delta_{n,m-1}$, with $n$ and $m$ being integer numbers ($n \geq 1$, $m \geq 1$). The Lanczos vectors are obtained by using the Lanczos recursive relations~\cite{Saad:2003}:
\begin{gather}
    \liou_{\bq} |\mathsf{q}_{n}\rangle = |\mathsf{q}_{n+1}\rangle \beta_{n+1}+|\mathsf{q}_{n}\rangle \alpha_{n}+|\mathsf{q}_{n-1}\rangle \gamma_{n}, \label{chains1}\\
    \liou_{\bq}^\dagger |\mathsf{p}_n\rangle=|\mathsf{p}_{n+1}\rangle \gamma_{n+1}+|\mathsf{p}_{n}\rangle \alpha_{n}+|\mathsf{p}_{n-1}\rangle \beta_{n}, 
    \label{chains2}
\end{gather}
where $\{\alpha_n, \beta_n, \gamma_n\}$ is a set of Lanczos coefficients that are defined as: $\alpha_n = \big\langle\mathsf{p}_n \big| \liou_{\bq} \big|\mathsf{q}_n \big\rangle$, 
$\beta_{n+1} = \sqrt{\langle\mathsf{u}_p|\mathsf{u}_q\rangle}$, and  
$\gamma_{n+1} = \beta_{n+1} \mathrm{sign}[\langle\mathsf{u}_p|\mathsf{u}_q\rangle]$, where 
$|\mathsf{u}_q\rangle = |\mathsf{q}_{n+1}\rangle \beta_{n+1}$ and $|\mathsf{u}_p\rangle = |\mathsf{p}_{n+1}\rangle \gamma_{n+1}$~\cite{gorni_spin_2018}. By setting $|\mathsf{p}_1\rangle = |\mathsf{q}_1\rangle =| \mathsf{v}_{\alpha'}\rangle$, the resolvent of the Liouvillian in Eq.~\eqref{eq:resolvent} can be expressed as a continued fraction~\cite{walker_efficient_2006, walker_ultrasoft_2007}:
\begin{equation}
    \chi_{\alpha\alpha'}(\bq, \omega) = \frac{1}{\omega -\alpha_1+\beta_2 \, \frac{\displaystyle 1}{\displaystyle \omega-\alpha_2+\dots} \, \gamma_2}.
    \label{liouvillian_LL}
\end{equation}
In practice, $\chi_{\alpha\alpha'}(\bq, \omega)$ must be converged with respect to the number of Lanczos iterations $n$ when solving Eqs.~\eqref{chains1} and \eqref{chains2} recursively.

We report in Sec.~S2 in the SI the evolution of the average between the even and odd LL coefficients $\alpha_n$ and $\beta_n$ along the Lanczos chain (the coefficient $\gamma_n$ is essentially equal to $\beta_n$ and for conciseness we display only the latter). As it was reported in previous studies \cite{malcioglu_turbotddft_2011,rocca_turbo_2008,gorni_spin_2018}, the $\alpha_n$ coefficient is very small and oscillates around zero (when the batch rotation is performed, namely when time-reversal symmetry holds, $\alpha_n=0$ by construction~\cite{walker_efficient_2006}), while $\beta_n$ is approximately equal to the half of the kinetic-energy cutoff in the wavefunction expansion ($\approx40=80/2$ Ry). We mention that the number of LL iterations necessary to converge TDDFPT$+U$ calculations is about 7000 -- 8000, which is substantially smaller than the 16000 iterations needed to converge TDDFPT. We attribute this difference to the fact that the $+U$ correction widens the band gap (see Table \ref{tab:ground_state}). As a consequence, the energy of Stoner excitations is blue-shifted, and less electronic transitions contribute to the system's response at low energies (where magnon excitations occur), stabilizing the convergence of the Lanczos chains.

Importantly, these equations are independent of the frequency $\omega$, meaning that they need to be solved only once for each value of the transferred momentum $\bq$ and each  Cartesian $\alpha$-component of the external magnetic field. The frequency only comes into play in the post-processing step, using the Lanczos coefficients $\{ \alpha_n, \beta_n, \gamma_n \}$ to compute the spin susceptibility tensor according to Eq.~\eqref{liouvillian_LL}. A small constant Lorentzian broadening $\eta$ is added to the frequency $\omega \rightarrow \omega + i\eta$ to regularize the cases when the frequency of the perturbation resonates with electronic vertical transition processes in the system (see the left-hand sides of Eqs. (\ref{stern::std}) and (\ref{stern::rev})). Since only the transverse component of $\chi_{\alpha\alpha'}(\bq, \omega)$ is needed to compute magnons, just one or two Lanczos chains are sufficient in some systems, depending on the system's symmetry and the direction of the ground-state magnetization.

\subsection*{Computational details}

All calculations are performed using the \QE\ distribution~\cite{giannozzi_quantum_2009, giannozzi_advanced_2017, Giannozzi:2020}. The ground-state calculations are carried out with the \textsc{PW} code~\cite{giannozzi_quantum_2009} using LSDA~\cite{perdew_accurate_1992} as the base xc functional. Optimized norm-conserving scalar-relativistic pseudopotentials~\cite{hamann_optimized_2013} are taken from the \textsc{PseudoDojo} library~\cite{van_setten_pseudodojo_2018}. We use a 80~Ry kinetic-energy cutoff for the plane-wave expansion of the KS wavefunctions and a 320~Ry cutoff for the charge density. The BZ is sampled with a $\Gamma$-centered $12 \times 12 \times 12$ $\bm{k}$-points grid. The spin-orbit coupling is neglected.

The Hubbard $U$ parameters are computed using DFPT~\cite{timrov_hubbard_2018, binci_noncollinear_2023} as implemented in the \textsc{HP} code~\cite{timrov_hp_2022}, with L\"owdin-orthogonalized atomic orbitals for Hubbard projectors~\cite{Lowdin:1950}. We employ uniform $\Gamma$-centered $\bm{k}$- and $\bm{q}$-point meshes of size $8 \times 8 \times 8$ and $4 \times 4 \times 4$, respectively, and use kinetic-energy cutoffs of 90~Ry for the KS wavefunctions and 360~Ry for the charge density, providing an accuracy for the Hubbard parameters of $\sim 0.01$~eV. The $U$ parameters are computed iteratively in a self-consistent manner as described in Refs.~\citenum{cococcioni_energetics_2019, timrov_self-consistent_2021}, which includes Hubbard forces and stresses in DFT+$U$ structural optimizations~\cite{timrov_pulay_2020}.

The magnon energies are computed using TDDFPT+$U$ and the LL approach, as implemented in a modified version of the \textsc{turboMAGNON} code~\cite{gorni_turbomagnon_2022}. We use ALSDA, both with and without Hubbard $U$. The TDDFPT and TDDFPT+$U$ calculations are performed at their respective optimized rhombohedral lattice parameters reported in Table~\ref{tab:ground_state}. The calculations employ the pseudo-Hermitian flavor of the Lanczos recursive algorithm~\cite{gorni_first-principles_2023, delugas_magnon-phonon_2023}, which includes an extrapolation technique for the Lanczos coefficients~\cite{Rocca:2008}. A Lorentzian smearing with a broadening parameter of 0.5 meV is used to plot the magnetic excitation spectra. All calculations are performed without symmetries since these are not yet implemented.



\section*{DATA AVAILABILITY}

The data used to produce the results of this work is available in the Materials Cloud Archive~\cite{MaterialsCloudArchive2025}. The \textsc{turboMAGNON} code including Hubbard corrections is part of a customized version of \QE which will be made publicly available with the official future releases.


\section*{ACKNOWLEDGEMENTS}

We thank Tommaso Gorni, Flaviano dos Santos, and Francesco Mauri for fruitful discussions.
We acknowledge support by the NCCR MARVEL, a National Centre of Competence in Research, funded by the Swiss National Science Foundation (Grant number 205602). L.B. acknowledges the Fellowship from the EPFL QSE Center ``Many-body neural simulations of quantum materials'' (Grant number 10060). This work was supported by a grant from the Swiss National Supercomputing Centre (CSCS) under project ID~s1073 and mr33 (Piz Daint).


\section*{Author contributions statement}
L.B. derived the equations and implemented the novel approach in \QE using preliminary I.T.'s routines. L.B. and I.T. tested the implementation and applied it to NiO and MnO. L.B., N.M. and I.T. analyzed the results and equally contributed to writing the manuscript.

\section*{Competing interests}

The authors declare no competing interests.

\bibliography{references, references2}

\begin{thebibliography}{100}
\bibitem{Lounisnote}
\bibinfo{note}{{To clarify potential misunderstandings, we note that the TDDFT method described in Refs.~\citenum{Lounis:2010, Lounis:2011} does not include the Hubbard $U$ correction from the DFT+$U$ framework. In these references, the letter ``$U$'' refers to the exchange-correlation kernel of the Dyson equation, not the ``Hubbard $U$'' parameter. The TDDFT method in Refs.~\citenum{Lounis:2010, Lounis:2011} is based on LSDA, and should not be confused with TDDFT+$U$.}}
\bibitem{Factor2}
\bibinfo{note}{{Note that in the equations for $J_{11}(\bm{q})$ and $J_{12}(\bm{q})$, there are double sums over the Cartesian indices ($\alpha$ and $\alpha'$), which results in each term appearing twice. For example, $\sum_{\alpha \neq \alpha'} \cos\pi(q_\alpha+q_{\alpha'}) = 2\cos\pi(q_x+q_y) + 2\cos\pi(q_x+q_z) + 2\cos\pi(q_y+q_z)$.}}
\end{thebibliography}

\end{document}


\maketitle

\section{Crystal and electronic structure properties of NiO and MnO computed using different functionals}

\begin{table}[h!]
    \centering
    \begin{tabular}{cllllll}
        \toprule
                             & Method    & $U$ (eV) & $a$ (\AA) & $\vartheta$ (deg) & $|\mM|$ ($\mu_\mathrm{B}$) & $E_g$ (eV) \\
        \midrule
        \multirow{5}{*}{NiO} & LSDA+$U$  &   6.26   & 5.03                 & 33.63              & 1.60                   & 3.04     \\
                             & PBE+$U$   &   5.42   & 5.19                 & 33.60              & 1.66                   & 2.87     \\
                             & PBEsol+$U$&   5.77   & 5.10                 & 33.62              & 1.64                   & 3.12     \\
                             & Expt.     &          & 5.11$^\mathrm{a}$    & 33.56$^\mathrm{a}$ & 1.77$^\mathrm{c}$      & 4.0$^\mathrm{e}$  \\
                             &           &          &                      &                    & 1.90$^\mathrm{d}$      & 4.3$^\mathrm{f}$  \\
        \midrule
        \multirow{5}{*}{MnO} & LSDA+$U$  &   4.29   & 5.32                 & 34.16              & 4.19                   & 1.93            \\
                             & PBE+$U$   &   3.85   & 5.46                 & 34.12              & 4.28                   & 1.64            \\
                             & PBEsol+$U$&   4.05   & 5.38                 & 34.19              & 4.23                   & 1.70            \\
                             & Expt.     &          & 5.44$^\mathrm{b}$    & 33.56$^\mathrm{b}$ & 4.79$^\mathrm{c}$      & 4.1$^\mathrm{e}$         \\
                             &           &          &                      &                    & 4.58$^\mathrm{d}$      & 3.9$\pm$0.4$^\mathrm{g}$ \\
        \bottomrule
    \end{tabular} 
    \caption{\textbf{Crystal and electronic structure properties of NiO and MnO}. Computed ground state properties using LSDA+$U$, PBE+$U$, and PBEsol+$U$ (all spin-polarized) with their respective self-consistent Hubbard $U$ parameters for Ni-$3d$ and Mn-$3d$ states calculated using DFPT, and as measured in experiments. The equilibrium rhombohedral lattice parameter ($a$), rhombohedral angle ($\vartheta$), magnetic moment ($|\mM|$), and band gap ($E_g$) are presented. The experimental values for $a$ and $\vartheta$ are determined from the cubic lattice using the experimental lattice parameter (4.17 and 4.43~\AA\ for NiO$^\mathrm{a}$ and MnO,$^\mathrm{b}$ respectively), since experimentally the rhombohedral distortion is not quantified. The angle $\vartheta = 33.56^\circ$ corresponds to the case with no rhombohedral distortion. Ref.$^\mathrm{a}$:~\citenum{Schmahl:1964}, Ref.$^\mathrm{b}$:~\citenum{Sasaki:1980}, Ref.$^\mathrm{c}$:~\citenum{Fender:1968}, Ref.$^\mathrm{d}$:~\citenum{Cheetham:1983}, Ref.$^\mathrm{e}$:~\citenum{Kurmaev:2008}, Ref.$^\mathrm{f}$:~\citenum{Sawatzky:1984}, Ref.$^\mathrm{g}$:~\citenum{vanElp:1991}.}
    \label{tab:ground_state}
\end{table}

\newpage

\section{Behavior of Lanczos coefficients}

\begin{figure}[h!]
\centering
\label{fig:LL_coeff}
\end{figure} 
%


\section{The $\mathbf{q} \rightarrow \mathbf{0}$ limit of the magnon dispersion}

\begin{figure}[h!]
\centering
\label{fig::q0limit}
\end{figure}

\bibliography{references, references2}

\section*{Figure Legends}
\textbf{Figure 1}. \textbf{Lanczos coefficients of NiO/MnO}. Behavior of the average even/odd Lanczos coefficients $\alpha_n$ (panels \textbf{a} and \textbf{c}) and $\beta_n$ (panels \textbf{b} and \textbf{d}) as a function of the number of Lanczos iterations $n$ within TDDFPT$+U$ for NiO and MnO.\\

\noindent
\textbf{Figure 2}. \textbf{Long-wavelenght behavior}. Magnon dispersion for NiO (a) and MnO (b) for small transferred momenta $\bq$ along the [111] direction ($\Gamma$-$X$) and a linear fit.